7th International Conference on Industry of the Future and Smart Manufacturing

# AI Asset Management for Manufacturing (AIM4M): Development of a Process Model for Operationalization

Lukas Rauh[a,]*, Mel-Rick Süner[a], Daniel Schel[a], Thomas Bauernhansl[a,b]

[a]*Fraunhofer Institute for Manufacturing Engineering and Automation IPA, Nobelstr. 12, Suttgart 70569, Germany*
[b]*Institute of Industrial Manufacturing and Management IFF, Allmandring 35, Stuttgart 70569, Germany*

**Abstract**

The benefits of adopting artificial intelligence (AI) in manufacturing are undeniable. However, operationalizing AI beyond the prototype, especially when involved with cyber-physical production systems (CPPS), remains a significant challenge due to the technical system complexity, a lack of implementation standards and fragmented organizational processes. To this end, this paper proposes a new process model for the lifecycle management of AI assets designed to address challenges in manufacturing and facilitate effective operationalization throughout the entire AI lifecycle. The process model, as a theoretical contribution, builds on machine learning operations (MLOps) principles and refines three aspects to address the domain-specific requirements from the CPPS context. As a result, the proposed process model aims to support organizations in practice to systematically develop, deploy and manage AI assets across their full lifecycle while aligning with CPPS-specific constraints and regulatory demands.

## 1. Introduction

In manufacturing, digital transformation combined with artificial intelligence (AI) is a key driver for efficiency and innovation [1]. As production systems grow in complexity and face volatile demand, cyber-physical systems (CPS) integrate machines and sensors with digital services for monitoring, optimization, and predictive decision-making. In the manufacturing context those CPS evolve into cyber-physical production systems (CPPS) [2]. The availability of industrial data, computational power, and storage has made integrating AI, especially machine learning (ML), into CPPS feasible. ML enables real-time analysis and adaptive control, improving efficiency [3]. Yet CPPS are fully connected systems of systems [4], and their platform constraints and interdisciplinary processes make adherence to safety standards and emerging requirements for privacy, security, and explainability essential.

* Corresponding author. Tel.: +49 711 970-1749
  E-mail address: lukas.rauh@ipa.fraunhofer.de

Against this backdrop, this work aligns the CPPS lifecycle with machine learning operations (MLOps) as the de-facto framework for reliable AI deployment. Building on this alignment, this work introduces AIM4M, a domain-specific extension of MLOps process models refined with role definitions, CPPS-specific activities, and compliance mechanisms. AIM4M explicitly addresses phases that are often underrepresented in traditional AI projects: governed operationalization, live updates during operation, and decommissioning of AI solutions (end of life), thereby closing critical gaps between prototypes and production.

Overview of this work: In Chapter 2 we synthesize the relevant background, derive neutral evaluation criteria (EC) and assess existing process models, via a systematic literature review (SLR), to finally identify gaps that we translate into design requirements (DR). Guided by those DRs and our methodological research approach (Chapter 3), we then present the AIM4M model as our main contribution and present its refinements (Chapter 4), followed by a concise discussion of implications, limitations, future work (Chapter 5) and the conclusion (Chapter 6).

## 2. State of the Art and gaps

This chapter outlines the background and specific challenges of applying MLOps in CPPS environments (Section 2.1), defines evaluation criteria (Section 2.2), presents findings and gaps from the SLR (Section 2.3), and concludes with the derivation of design requirements (Section 2.4) as the basis for the AIM4M process model.

*2.1. Background*

The growing adoption of AI systems in manufacturing brings a rising demand to manage their entire lifecycle in a structured and systematic way. A typical AI lifecycle comprises multiple iterative phases connected by feedback loops, ranging from data preparation and modeling through evaluation and integration to operational deployment and continuous improvement [4], [5]. Established approaches such as CRISP-ML(Q) extend classical models by explicitly integrating quality assurance (via QA checkpoints) and feedback mechanisms [6]. While those approaches provide solid foundations, current MLOps process models often remain at high level and under-specify steps and artifacts [7].

Additionally, the distribution of roles within the lifecycle becomes increasingly significant. Research emphasizes that successful industrial AI initiatives rely on cross-functional collaboration involving not just AI specialists but also domain stakeholders, data scientists, and software or infrastructure engineers [4], [5], [6]. Complementary perspectives, such as those in [8], [9], identify further MLOps-related responsibilities that call for the involvement of hardware engineers, infrastructure architects, and quality assurance professionals: roles that are still insufficiently defined in most current lifecycle models. These observations motivate our evaluation criteria EC5 (live model updates through feedback), EC7 (governance and QA) and our criteria for systematic lifecycle engineering: EC1 (roles and responsibilities per step), EC2 (detailed, stepwise process engineering), and EC3 (comprehensive lifecycle management from ideation to end-of-life).

Furthermore, to also take CPPS specifics into account, life cycle models for AI in the MLOps approach must consider not only software components (e.g., data pipelines and versioning) but also operational activities in a production-related hardware context. However, while current approaches increasingly provide important guidance for practice-oriented lifecycle design, the methodical integration of such physical production environments, especially for systems with hard constraints by the operational technology (OT) environment, into established life cycle models has so far been largely ignored [10]. Moreover, the use of AI in production poses not only technical but also regulatory challenges. With the EU AI Act, there is an increasing focus on requirements aimed at transparency, traceability, and documentation in the AI lifecycle [11]. One concept that can contribute to the structured implementation of these requirements is the use of so-called AI cards, as defined in [12]. These standardized artifacts record key information on data, model behavior, limits, and risks, supporting internal QA and regulator-facing communication and enabling responsible AI operation, especially in CPPS. These observations motivate our domain specific evaluation criteria EC4 (CPPS for industry 4.0) and our regulatory criteria EC6 (regulatory compliance).

## 2.2. Evaluation criteria for State-of-the-Art Analysis

Based on the background and mentioned challenges in Section 2.1 and our industrial experience, we derive seven neutral criteria to assess the suitability of process models for AI lifecycle management in CPPS. These criteria measure coverage (not prescribe solutions) and form the basis for review and following gap analysis of the state of the art.

Table 1. Evaluation criteria dimensions, based on distilled challenges, clustered by high-level categories.

| Category | Evaluation Criteria Dimensions |
| --- | --- |
| Systematic lifecycle engineering | (EC1) Responsibilities and actors: Relevant roles and responsibilities are listed per step/phase in the process. |
| | (EC2) Systematical engineering: Process steps are defined and described in detail instead of a broad or generic model. |
| | (EC3) Comprehensive lifecycle management (LCM): The process is detailed from idea to end-of-life. |
| Domain-specific adaptation | (EC4) CPS/CPPS for Industry 4.0 context: The process integrates AI artifacts in the CPPS context or their lifecycle. |
| | (EC5) Live model updates: The process allows continuous optimization with model iterations during the operation phase. |
| Regulatory requirements | (EC6) Regulatory compliance: Challenges (safety, explainability, etc.) are known. Assessments are involved. |
| | (EC7) Governance and quality assessment (QA): Challenges are known, and assessments are involved. |

We apply EC1-EC7 in Section 2.3 using a Harvey-Balls scale to compare existing models as the state of the art. The scoring rubric and methodological details are summarized in Chapter 3.

## 2.3. SLR Results and Gap Analysis

Using the PRISMA framework we screened 275 records (243 database hits plus 32 prior sources), de-duplicated to 196, and, after applying inclusion/exclusion criteria, retained 11 primary studies for assessment (method details in Chapter 3). Using EC1-EC7, we evaluated each study with Harvey-Balls (0/50/100% coverage) and computed an average score per study. Fig. 1 summarizes the results: scores range from 21% to 64% (mean under 50%). Most models address lifecycle coverage including governed live updates (EC3, EC5, partly EC1) and quality governance (EC7), but show lower coverage in CPPS integration under OT/IT constraints (EC4), and regulatory compliance (EC6) with basic systematic structuring (EC2).

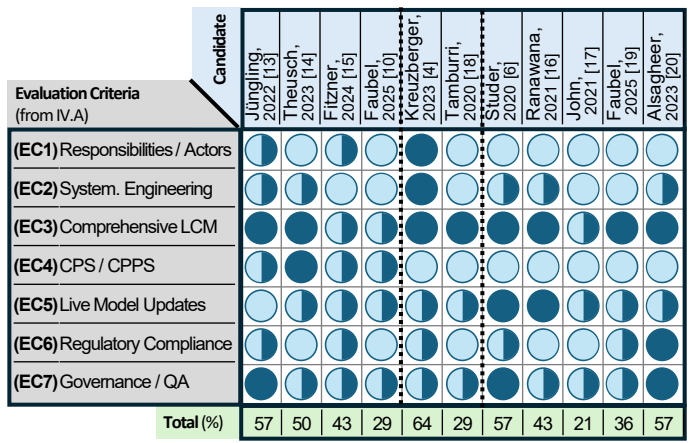

Fig. 1. Assessment of the 11 primary studies ([4], [6], [10], [13], [14], [15], [16], [17], [18], [19], [20]) for MLOps process models, based on the set of evaluation criteria (see Table 1).

The process models cluster into three patterns (dotted outlines in Fig. 1). First, CPPS-explicit approaches (e.g., Theusch) acknowledge OT/IT and shop-floor realities but remain incomplete on roles, systematic engineering, or compliance. Second, generic foundations (e.g., Kreuzberger) achieve the highest breadth (up to 64%) yet lack depth in live updates, compliance, and governance. Third, specialized models (e.g., Studer for QA, Alsagheer for governance, Ranawana for agile lifecycles, Faubel for explainability, John for maturity levels) offer strong niche coverage but remain at or below mid-level overall. No single model covers all seven criteria simultaneously. These consistent gaps motivate the design requirements in Section 2.4 and, ultimately, the AIM4M model.

*2.4. Design requirements*

Building on the SLR findings and identified gaps in Section 2.3, we derive seven design requirements (DR1-DR7) that a CPPS-oriented MLOps process model must satisfy. Each DR maps directly to an evaluation criterion (EC1-EC7) and is stated as a prescriptive "must" for the process model design.

- DR1: Responsibilities and roles (EC1): The model must define domain-specific roles across OT/IT, quality, and compliance, and assigns clear responsibilities (e.g., lead and contributor roles) to process stages.
- DR2: Systematic engineering (EC2): The model must prescribe defined stages with inputs and deliverables, process steps, and quality gates that enable consistent progression and auditability.
- DR3: Comprehensive lifecycle management (EC3): The model must cover the end-to-end lifecycle from ideation to end-of-life, for data, models, and deployments.
- DR4: CPPS for Industry 4.0 context (EC4): The model must integrate CPPS-specific activities, including OT/IT integration or deployment design, and hybrid testing in the lab and shop-floor prior to release.
- DR5: Live model updates (EC5): The model must support governed updates in operation with continuous monitoring, drift detection, automated (re)training and redeployment via reusable pipelines, controlled rollout (e.g., canary), and rollback procedures, gated by approvals. This is intended to keep compliant with the evolving environment in regulated context, as often in manufacturing.
- DR6: Regulatory compliance (EC6): The model must embed compliance mechanisms with documented checks and approvals, structured documentation artefacts, aligned with applicable regulations and standards.
- DR7: Governance and quality assessment (EC7): The model must define lifecycle-wide QA and governance mechanisms, including practices like governed decision gates to ensure transparency and accountability.

These design requirements provide the acceptance criteria for the AIM4M artefact. In Chapter 4, AIM4M is presented as the result of the design process to fulfil DR1-DR7, and traceability from each design requirement to specific model elements is demonstrated in Section 4.3.

## 3. Methodology

Developing a process model for AI asset management in manufacturing requires a methodologically sound development approach. To this end, this work follows the design science research (DSR) methodology to derive our main contribution, the AIM4M process model, as a problem-solving artifact grounded in scientific evidence and practice. Fig. 2 provides a high-level map of how the development approach is structured across the DSR cycles and indicates where each step is covered in this work.

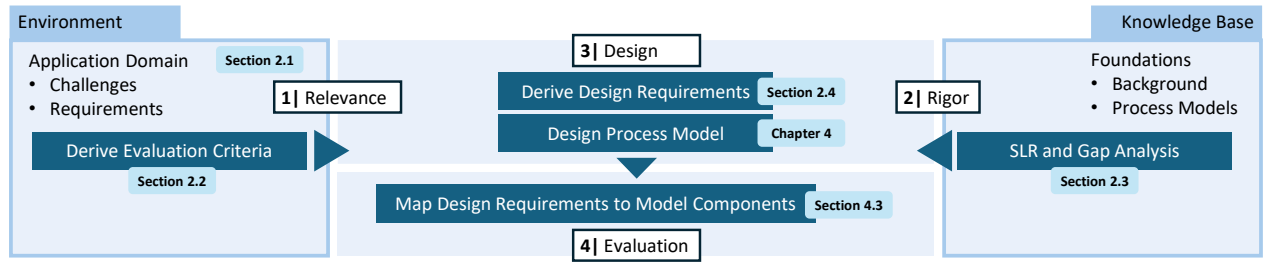

Fig. 2. Methodology overview of the development approach for this work, mapped to the DSR with its structure and cycles. Section and Chapter labels in light blue indicate where each step appears in this work.

The following list maps the development approach onto the DSR cycles:
1. **Evaluation criteria (relevance cycle)**: We frame the problem in the CPPS context and synthesize background on MLOps and regulation (Section 2.1) to derive seven neutral evaluation criteria EC1-EC7 (Section 2.2). These criteria capture three challenge clusters: systematic lifecycle engineering, domain-specific CPPS adaptation, and regulatory/governance. These criteria guide the subsequent SLR and gap analysis (rigor cycle) and serve as input to artefact design (design cycle).

2. **Gap analysis (rigor cycle)**: We investigate the knowledge base of this cycle via a PRISMA-guided SLR [21]. Search was conducted in January 2025 on Scopus and Web of Science (title/abstract/keywords). After de-duplication and screening, 11 primary studies remained (search string, screening criteria, and PRISMA flow are summarized below in Fig. 3). Each remaining study then has been qualitatively assessed against EC1-EC7 using a three-level Harvey-Balls scheme (0/50/100% coverage), with per-study averages to aid comparability. The degree of coverage or fulfilment per EC refers to the qualitative assessment of whether the criterion is met, for example through addressed requirements or identifiable keywords in the process models. The assessment and the resulting gap analysis are reported in Section 2.3.
3. **Design (design cycle)**: We translate the observed gaps into prescriptive design requirements DR1-DR7 (Section 2.4) and use them to guide the design of AIM4M. The model integrates strengths from the best-performing approaches (knowledge base) while adding CPPS-specific activities, role/governance refinements, and compliance mechanisms (application domain). AIM4M was iteratively refined with stakeholder feedback to ensure practical relevance and tool-agnostic applicability and is presented as our main contribution in Chapter 4.
4. **Evaluation (evaluation cycle)**: Following the DSR methodology, the evaluation of the AIM4M process model as an artifact involves a systematic evaluation based on the design requirements DR1-DR7 from the design process, as a result synthesized from environment and knowledge base. Additionally, and in line with DSR evaluation methods, to follow up on this work with its artifact, a practical validation via an empirical case study with experts will follow in our future work and accompanying papers.

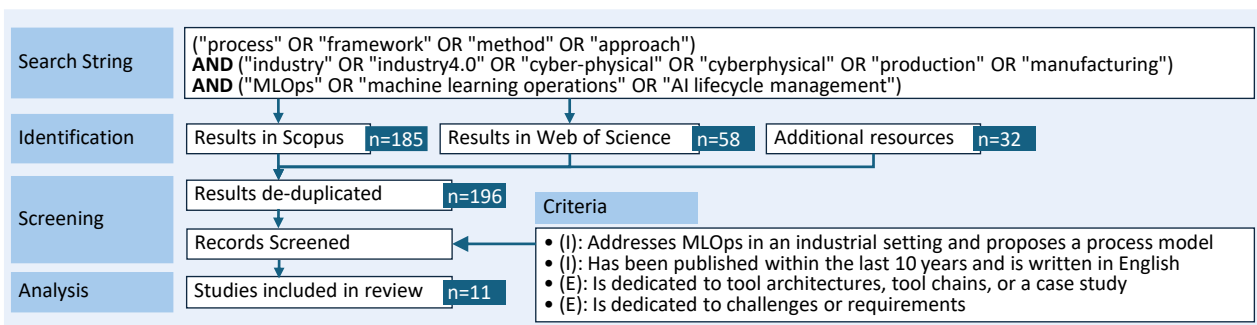

Fig. 3. PRISMA flow, including search string and screening criteria, conducted in January 2025.

## 4. The AIM4M Process Model

This Chapter introduces the AIM4M process model as the result of the development process, as outlined in the design cycle in Chapter 3. It represents the main contribution of this work as a CPPS-oriented MLOps process model that builds on the typical process definition of MLOps as a framework for AI lifecycle management (Section 2.1), with a refinement based on the identified gaps in recent process models (Section 2.3) and finally operationalizing the design requirements (DR1-DR7), derived in Section 2.4. The model spans the full lifecycle with four phases (instead of typically three phases) and introduces three domain-specific refinements: extended OT/IT roles, CPPS activities for OT/IT integration, and embedded compliance mechanisms. It serves as a practical artifact to guide organizations in integrating AI into the CPPS product lifecycle, independent of their specific technology stack or tools. Section 4.1 summarizes the phases, Section 4.2 details the refinements (roles, CPPS activities, compliance), and Section 4.3 maps each DR to concrete AIM4M elements to demonstrate coverage.

*4.1. Overview and Lifecycle Phases*

Building on the overview above, this subsection summarizes the structure of AIM4M and how it organizes the AI lifecycle under CPPS constraints. As shown in Fig. 4 AIM4M partitions the whole lifecycle into four phases (ideation, development, operation and retirement) and eleven stages (I-XI, roman numerals). Each stage groups several steps and activities (arabic numerals, purple marker mark refined CPPS activities; see Section 4.2) with defined inputs and outputs, is connected by feedback loops, and is governed by approval gates at phase transitions, that support decision-making and enforce structured documentation. The required roles are specified for each stage (purple marks the refined roles; see Section 4.2). Across all stages, artefact traceability is ensured via versioning, and structured AI Cards (use case, dataset, model, deployment) are produced as key outputs at designated stages. To aid readability, phases and stages are color-coded: ideation appears in light blue, development in dark blue, operation in mint, end-of-life in black, and quality gates are shown in yellow.

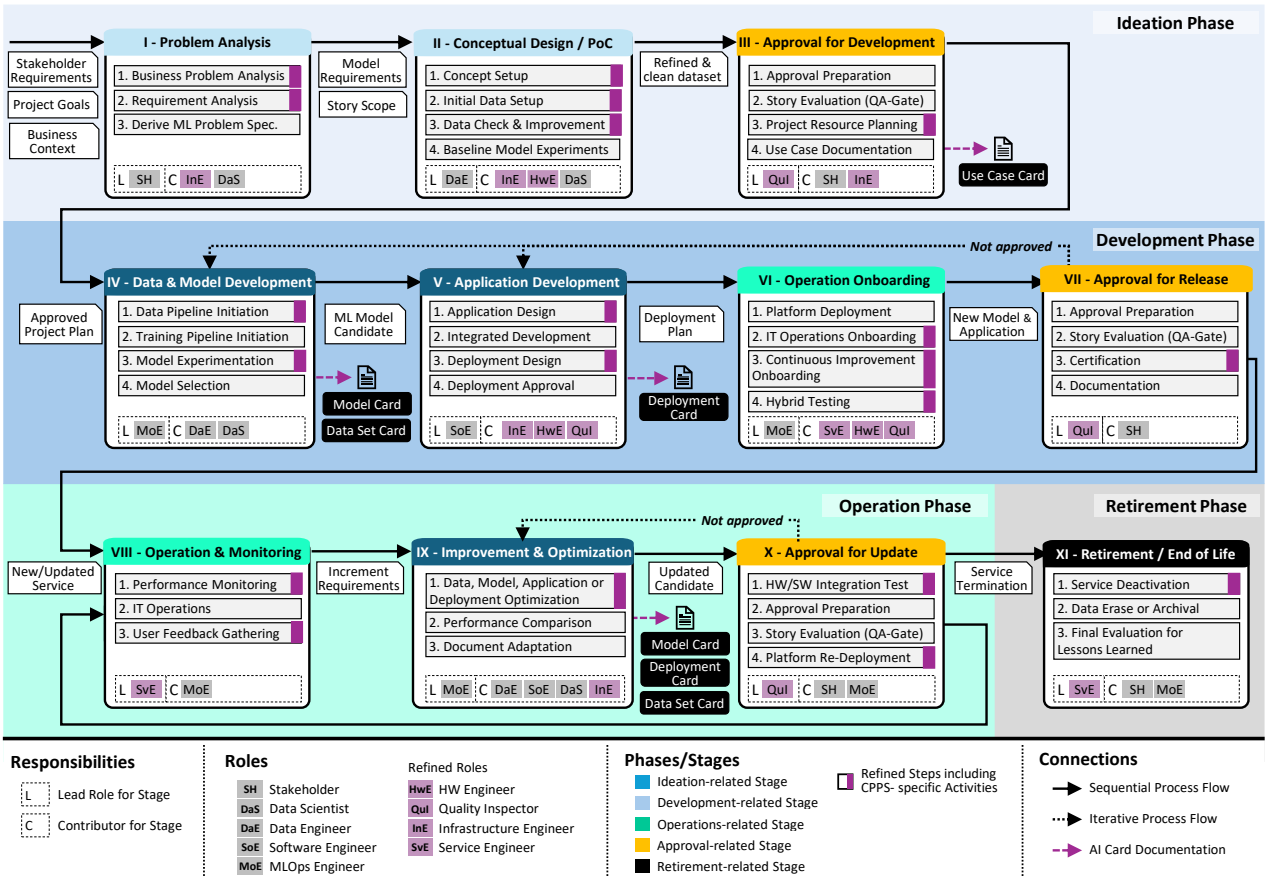

Fig. 4. The proposed AIM4M process model, with domain-specific refinements (purple marked roles, process steps, and AI cards).

It follows a brief description of each individual phase, including the quality gates:

- The **ideation phase** (stages I to III) defines and validates new AI initiatives/prototypes, aligning them with business goals, stakeholder commitment, and technical feasibility. Data availability and CPPS boundaries are validated. Passing the first gate (approval for development) unlocks development resources and transitions from prototype-level to full-scale development.
- The **development phase** (stages IV to VII) focuses on technical realization: building data and model pipelines, designing application and deployment architecture, and preparing for operational handover into existing infrastructure. Passing the second gate (approval for release) authorizes the initial release of the AI solution with the embedded AI model to customers.
- The **operation phase** (stages VIII to X) focuses on continuous optimization, model re-evaluation, and consistent operational management. Iterative cycles of improvement and redeployment reflect the dynamic nature of MLOps practices, ensuring adaptability to evolving requirements. Documentation and AI Cards are also consequentially and iteratively updated.
- The **retirement phase** (stage XI) addresses de-commissioning, ensuring responsible management of end-of-life applications through archiving and data and asset handling executed under policy and regulation.

*4.2. Refinements*

Compared to existing MLOps models and based on the identified gaps in Section 2.3, AIM4M introduces three domain-specific refinements to address CPPS constraints and regulatory demands: (i) adopted role model to match the OT/IT context; (ii) CPPS activities embedded across all phases; and (iii) compliance mechanisms with structured artefacts and approvals. In the illustration of the model, these refinements are highlighted in purple (see Fig. 4).

(i) With the refinement of the scope from general MLOps to the manufacturing domain, we extend the general scope of relevant roles for an AI lifecycle with four additional roles. These additional roles mainly originate from the shift from pure software systems to an OT/IT converging cyber-physical system. This introduces the development of hardware, complex solution designs, intensified hybrid testing, and, finally, the deployment and operation within a production system. Additionally, more roles originate from introducing compliance mechanisms into the process model. This introduces roles with elevated responsibilities and roles with auditor activities. The following Table 2 lists additional roles and their activities.

Table 2. Additional roles in the AIM4M process model, representing OT and QA specifics.

| Role | Related Activities and Responsibilities |
|---|---|
| Hardware Engineer (HwE) | <ul><li>Specify and validate machinery, sensor, and IIoT device integration for data acquisition and inference</li><li>Ensure compatibility between AI applications and machine-level hardware systems</li></ul> |
| Infrastructure Engineer (InE) | <ul><li>Design comprehensive (edge, hybrid, cloud) solution architecture for the AI solution</li><li>Ensure sufficient infrastructure resources to run and scale all relevant components</li></ul> |
| Service Engineer (SvE) | <ul><li>Monitor and maintain deployed AI systems in operational environments</li><li>Report anomalies and support root cause analysis for model behavior in the field</li></ul> |
| Quality Inspector (QuI) | <ul><li>Define and execute validation protocols for AI-assisted decision outputs</li><li>Verify model results against product quality specifications and safety thresholds</li></ul> |

(ii) AIM4M integrates CPPS activities across the lifecycle. In ideation (I-III), teams define system boundaries and AI-asset interfaces, align OT/IT concepts, and establish an initial IIoT data setup to verify availability and quality, informing resource planning for hardware integration. In development (IV-VII), real sensor and machine data feed reproducible data/model pipelines and comprehensive test sets (including edge cases), application and deployment are designed for interoperability and low-latency edge execution. Then the responsible roles conduct hybrid testing in lab and pilot environments and plan certification-relevant documentation. In operation (VIII-X), real-time monitoring and user feedback drive governed retraining and controlled rollouts via reusable pipelines. At end-of-life (XI), responsible physical-asset and data retention are applied, consistent with policy and regulation.

(iii) Compliance mechanisms are embedded throughout the AIM4M process model. Structured AI Cards (covering the use case, dataset, model, and deployment) capture relevant information at key stages (III, IV, V, IX) and are iteratively updated as the project evolves. Quality gates (III, VII, X) enforce documented approvals and require segregation of duties to ensure governance. Within these gates, verification and validation activities (risk and conformity checks) are supported to align with regulations and standards. Together this creates an end-to-end audit trail, spanning data, models, and deployment artifacts to enforce transparency, reproducibility, and accountability under the specific constraints of CPPS environments.

*4.3. Traceability to Design Requirements*

This section links AIM4M to the defined design requirements (DR1-DR7) that are derived from the evaluation criteria (EC1-EC7) discussed in Sections 2.2 and 2.4. Table 3 below provides an illustrative mapping matrix from design requirements to specific realized model elements. This traceability demonstrates that AIM4M satisfies the design requirements while preserving a rigorous yet adoptable lifecycle structure. As Fig. 4 illustrates, the integration of roles, CPPS activities, and compliance mechanisms operationalizes these design requirements without prescribing specific technologies, thereby maintaining the intended technology-agnostic character of the model.

Table 3. Mapping between design requirements and process model elements.

| ID | Requirement summary | Addressed in | Model elements (roles / artifacts / gates) |
|---|---|---|---|
| DR1 | Clear roles and responsibilities | All stages | Extended roles; Stage responsibilities; AI Cards ownership |
| DR2 | Systematic engineering with defined stages | All stages | Staged activities with steps; sequenced flow with feedback loops |
| DR3 | End-to-end lifecycle coverage incl. EoL | All stages | Inclusion of retirement/EoL (XI) and archival process steps |
| DR4 | CPPS integration (OT/IT, edge/hybrid) | All stages | CPPS activities as process steps, e.g. hybrid testing or edge deploy |
| DR5 | Safe live model updates in operation | VIII-X | Process steps for pipelines and update gate (X); AI Card revisions |
| DR6 | Embedded compliance and documentation | III, VII, X | Gate tasks; AI Cards; evidence archives |
| DR7 | Quality assurance | III, VII, X | Quality gates and monitoring based on defined requirements |

## 5. Discussion

This chapter discusses the applicability, limitations, and potential impact of the proposed AIM4M process model for MLOps-based AI lifecycle management in CPPS. As a theoretical artifact developed through DSR, the proposed AIM4M process model is intended to guide stakeholders across the full AI lifecycle with a core process logic while subsequent work will address its implementation and validation through case studies and software tooling.

### 5.1. Applicability

AIM4M targets AI adoption in CPPS environments and operates at the process level rather than prescribing a specific technology stack or implementation framework. It is broadly applicable for AI development in manufacturing across sectors (e.g., automotive, aerospace, pharmaceutical) and geographies. While the model is aligned first and foremost with European regulatory expectations, its lifecycle abstraction and dynamic documentation approach (e.g., quality gates supported by AI-Cards) enable transferability to other regulatory contexts.

Applicability is given where at least the commitment to structured AI development has been established. There are no data availability and maturity requirements, since the model has no direct impact on the modelling process itself, and instead operates at a higher level of lifecycle management and administration. The model's modular process architecture with the generic process core supports tailoring to production realities and maturity levels. Rather than prescribing a one-size-fits-all model with prescribing nature, AIM4M is intended to be applied as a flexible, context-sensitive process model, where staged adoption is also supported, when roles or responsibilities are bundled to match resource constraints. This scalability is achieved by adjusting automation depth and role granularity: small and medium-sized enterprises (SMEs) can bundle responsibilities while preserving governance, whereas large enterprises can specialize and automate for high-throughput, compliance-driven operation.

The intended audience comprises industrial practitioners and decision-makers across roles in development, deployment, and governance of AI in manufacturing (cf. legend in Fig. 4). At the management level (IT/production managers), it provides role clarity, a governance structure, and decision gates that support strategic planning and assignment of responsibilities. At the technical level (ML/data/IT engineers, quality officers), it offers detailed process steps, iteration paths, and documentation scaffolding for teams building and operating AI systems.

### 5.2. Limitations and Future Work

The presented evaluation is ex-ante and analytical in the DSR tradition, mapping design requirements to model features, which strengthens transparency but limits the validation scope at this stage. The consolidated requirement set, while grounded in literature and practice, is selective. Hence, future work should include field evaluations and longitudinal studies.

By design, AIM4M operates at the process level and remains technology-agnostic, enhancing portability across tech stacks, sectors, and regulations, yet it deliberately stops short of prescribing instantiation artifacts such as deployment blueprints or audit templates. The core process lifecycle is intended to remain stable while the content of gates and documentation can be updated as regulations and norms evolve. Future work should include iterative design updates and options for explicit instantiation artifacts.

The effectiveness of AIM4M is dependent on the context. Outcomes will vary with manufacturing complexity, OT/IT infrastructure, and AI/IT readiness. Highly automated factories can benefit immediately, whereas organizations with fragmented data pipelines or limited MLOps resource capacity may face integration and change-management hurdles. As an example, compliance workflows and documentation introduce overhead, particularly in regulated settings already subject to extensive record-keeping. AIM4M addresses these frictions by providing role clarity, decision gates, and standardized yet lightweight documentation (e.g., AI-Cards) to reduce planning effort, documentation burden, and operational risk over time, though it cannot eliminate them entirely. Future refinements should aim to balance rigor and usability by tailoring documentation depth and automation support to maturity levels and risk profiles.

*5.3. Impact and Contribution*

This work contributes to AI lifecycle management for manufacturing along three axes: knowledge base, process model artefact, and practice. First, we systematize CPPS-specific needs by introducing neutral evaluation criteria (EC1-EC7), executing a PRISMA-guided SLR combined with a consistent Harvey-Balls scoring scheme, and synthesizing prescriptive design requirements (DR1-DR7). The gap analysis (Fig. 1, Section 2.3) shows that existing models, while adopting MLOps principles, lack a structured, domain-specific approach spanning the full product lifecycle. In the CPPS context, gaps are pronounced in regulatory compliance, validation in hybrid OT/IT testing environments, and robust AI asset management across product increments. By turning these gaps into explicit requirements, we provide an approach to close a methodological gap in the knowledge base where challenges are often listed but not operationalized into artefacts.

Second, we contribute AIM4M, a CPPS-oriented MLOps process model that embeds quality assurance throughout the lifecycle to maintain production constraints and safety-critical standards. AIM4M operationalizes DR1-DR7 (while no other model covers all seven EC; see Section 2.3) across four phases and eleven stages with explicit gates (before development, release, and each update), structured documentation (AI Cards for use case, dataset, model, deployment), and end-to-end audit readiness. It adds CPPS-specific activities and compliance mechanisms and is tool-agnostic to fit diverse tech stacks.

Third, in practice AIM4M helps organizations move beyond pilots to reliable operation. It addresses both management and technical audiences (Section 5.1): leadership gains a clear roadmap and approval strategy, while implementation teams receive concrete guidance and artefacts to execute under CPPS constraints. By aligning each phase with AI Cards and governed gates, AIM4M increases reproducibility and risk mitigation, keeps AI assets traceable and auditable, and guides compliant retirement through a formal end-of-life stage, closing the loop on regulatory alignment and operational transparency.

*5.4. Positioning and Expected Effects*

AIM4M is positioned as a domain-specific reference process model for organizations aiming to adopt AI responsibly in CPPS environments. Rather than functioning as a tooling framework, it provides a blueprint for aligning AI development and operations with organizational, regulatory, and technical needs. The expected effects include:
- An improved traceability and audit readiness, through structure documentation and systematic quality gates
- More predictable rollouts across heterogeneous environments, through including CPPS refinements

The process model scales from SME-friendly implementations with bundled roles to fully automated enterprise setups with fine-grained governance (Section 5.1). Future empirical studies should quantify these effects using operational KPIs such as model development time or audit effort. These studies will also help derive stack-specific instantiation patterns and assess AIM4M's adaptability across different manufacturing contexts (Section 5.2).

# 6. Conclusion

This work proposes the AIM4M process model for systematic artificial intelligence (AI) asset management tailored for cyber-physical production systems (CPPS), addressing practical challenges. It refines existing process models for machine learning operation (MLOps), following the design science research (DSR) methodology, and shifts from a general AI scope to a domain-specific approach that aligns with CPPS requirements to fulfill practical requirements and regulatory standards. Therefore, the AIM4M process model introduces refinements for domain-specific roles, CPPS activities, and compliance mechanisms to provide a tool-agnostic, governance-ready process model for AI products or AI applications in CPPS, from ideation to end-of-life.

Future work will focus on gaining more practical experience by applying AIM4M in more customer projects to accumulate empirical evidence and iteratively refining model elements, while preserving the core process logic. In parallel, the AIM4M process model will be accompanied with software tooling that implements roles, gates, and documentation through integrations with existing MLOps toolchains. Therefore, Industry 4.0 standards (e.g., OPC UA, Asset Administration Shell) will be incorporated to enable plug-and-produce interoperability for developed AI applications, including configurations suitable for organizations with lower AI maturity. These steps will lead to concrete instantiation patterns and templates, translating the theoretical process artifact into actionable practice.

## Acknowledgements

This work is supported by the project "AI Matters", co-funded by the European Union under Grant Agreement No. 101100707 and by the State of Baden-Württemberg, according to a decision of the State Parliament. However, the views and opinions expressed are those of the author(s) only and do not necessarily reflect those of the European Union or the European Commission. Neither the European Union nor the granting authority can be held responsible.